# Single-dot spectroscopy via elastic single-electron tunneling through a pair of coupled quantum dots

T. Ota[1], K. Ono[2], M. Stopa[1], T. Hatano[1], S. Tarucha[1,2,3],
H. Z. Song[4], Y. Nakata[4], T. Miyazawa[4], T. Ohshima[4], N. Yokoyama[4]

[1]*Mesoscopic Correlation Project, ERATO, JST, Atsugi-shi, Kanagawa 243-0198, Japan*
[2]*University of Tokyo, Bunkyo-ku, Tokyo, 113-0033, Japan*
[3]*NTT Basic Research Laboratories, Atsugi-shi, Kanagawa 243-0198, Japan*
[4]*Fujitsu Laboratories Ltd., Atsugi-shi, Kanagawa 243-0197, Japan*

We study the electronic structure of a *single* self-assembled InAs quantum dot by probing elastic single-electron tunneling through a single pair of weakly coupled dots. In the region below pinch-off voltage, the *non-linear* threshold voltage behavior provides electronic addition energies exactly as the linear, Coulomb blockade oscillation does. By analyzing it, we identify the s and p shell addition spectrum for up to six electrons in the single InAs dot, i.e. one of the coupled dots. The evolution of shell addition spectrum with magnetic field provides Fock-Darwin spectra of s and p shell.

Probes of the electronic structure of single, nanometer scale structures are currently in the forefront of experimental condensed matter physics. Innovative fabrication has combined with new optical and electronic spectroscopic techniques to examine the properties of individual artificial atoms and molecules [1], embedded impurities [2] and nanotubes [3] as well as other natural molecules [4]. In each case, the method by which the electronic properties are accessed is intrinsically entwined with the way that the nanoscale object is configured in the fabrication process.

Of particular interest, both for their immense potential for optical and electronic device applications and for their unique physical properties [5], are InAs quantum dots grown via the self-assembly technique. Recently, optical spectroscopy of individual InAs dots has been achieved [6]. Such investigations, however, necessarily involve the interaction of electron-hole pairs. In addition, manipulating the total charge state of the dot (e.g. producing charged excitons or trions) in a controllable fashion is challenging. By contrast, the complementary approach of studying the electronic structure through transport experiments has only recently shown progress, due to the difficulty of attaching leads to the dots. Initially, studies employing capacitance spectroscopy [7] were limited to large ensembles of self-assembled dots, which complicated the interpretation of the data due to intrinsic inhomogeneity in the samples. To date, only studies on the level structure of single, empty InAs dots embedded in vertical heterostructures [8] and the multi-electron regime of single dots embedded in a quantum point contact [9] have been reported.

In this paper, we study the electronic structure of a *single* InAs dot by probing it, via serial transport, with a second smaller dot located vertically above it [10]. We employ a double layer of InAs dots, embedded in a sub-micron size pillar, and isolate a single vertical pair of dots by depleting the structure with a side gate [11]. We observe Coulomb diamonds in the gate voltage, $V_G$, source-drain voltage, $V_{SD}$, plane which unambiguously confirm that we can deplete the double dot all the way down to total electron number $N_T=0$. We probe the electronic addition spectrum of the larger dot by measuring the threshold $V_{SD}$ dependence on $V_G$ below the Coulomb diamond regime, where the smaller dot electron number $N_S$ is zero and the larger dot electron number $N_L$ ranges from six down to zero. While the Coulomb diamonds show evidence of current contribution from neighboring dot pairs (evidenced in overlapping diamonds with differing slopes), by depleting into the sub-diamond regime we are able to obtain clean results on the last conducting dot pair in the pillar (*position-sensitive* single-electron tunneling spectroscopy technique) [11]. In the sub-diamond regime the *non-linear* threshold voltage behavior provides electronic addition energies exactly as the linear, Coulomb blockade oscillation structure does in the diamond regime. We are able to identify the s and p shell addition spectrum for up to six electrons in the larger dot. We study the evolution of this spectrum with a transverse magnetic field (B-field) and analyze it using the standard Fock-Darwin (F-D) spectrum, in a manner similar to that employed in the study of weakly coupled, vertical GaAs (not self-assembled) double dots [12].

The material that we use consists of a 750 nm-thick n-doped GaAs buffer layer on GaAs (100) semi-insulating substrate, a 36 nm i-GaAs layer incorporating two InAs self–assembled quantum dot layers in the center and a 400-nm thick n-doped GaAs layer. InAs self-assembled dots are formed on an InAs wetting layer, and the two wetting layers are 14.5 nm apart. The typical height and diameter of the dots are 4 nm and 40 nm, respectively. The average density of the dots is on the order of $10^{10}$ cm$^{-2}$. Using this material, we have fabricated circular mesa samples as shown in Figure 1(a). Details of the sample structures are described in Ref. 11. About six coupled InAs dots are embedded in the 0.25 µm-diameter circular mesa. Each tunnel barrier is sufficiently thick that electrons tunnel sequentially. Transport measurements are performed at 100 mK in a dilution refrigerator.

Although about six coupled dots exist in this pillar, we are usually able to observe only a *single* pair of the dots near the pinch-off point using a *position-sensitive* single-electron tunneling spectroscopy technique. When two or more pairs of coupled dots are located close together, different families of Coulomb diamonds, which are bounded by threshold lines with different slopes, are observed in the non-linear transport regime [11], indicating that the gate capacitance to each pair of coupled dots strongly depends on its location [8].

Figure 1(b) shows a color log-scale plot of $dI/dV$ as a function of $V_{SD}$ and $V_G$. In the region of $V_G > V_0 = -0.85$ V, a few closed Coulomb diamonds are observed along $V_{SD}=0$ V. On the other hand, in the region of $V_G < V_0$, irregular structures indicated by the dotted lines are observed in the non-linear transport regime [13]. "Kinks" for negative $V_{SD}$ and "vertical lines" for positive $V_{SD}$ are always observed in pairs, as indicated by the dot-dashed line. Usually, in a weakly coupled asymmetric dot system, these irregular structures in the sub-diamond region are observed near $V_0$ [14]. At the point $V_G =V_0$, we find that there are six electrons trapped in the larger dot and no electrons in the smaller dot, i.e. $(N_S, N_L)=(0, 6)$. With increasingly negative $V_G$, the electrons in the dots are reduced one by one, as each dash-dot line is crossed, from $(N_S, N_L)=(0, 6)$ to $(N_S, N_L)=(0, 2)$. In this paper, we focus on analyzing the "vertical lines" (VLs) observed along the conductance threshold in the positive $V_{SD}$ range.

The detailed structure of the VLs is shown in Fig. 2, including a wider positive $V_{SD}$ range. From $V_{SD} = 0$ to 200 mV, six VLs are observed along the conductance threshold, as indicated by the dotted lines. For negative $V_{SD}$, six "kinks," which pair with the six VLs, are indicated by the dot-dashed lines. (The VL at B (D) pairs with the kink at F (G) in the lower inset of Fig. 2.) Results from a simple constant interaction (CI) model calculation for the threshold voltage structure in the sub-diamond region, are shown in the upper inset of Fig. 2. The calculation takes charging and confinement energies as input parameters [15] (discussed below). Correspondences between the calculation and the observations allow us to assign energy level diagrams (Fig. 2, lower panels A-F) to various points in the $V_{SD}$-$V_G$ plane (labeled A-F in upper panel).

To explain the structure of the current threshold, we first note that the gate couples about equally to the two dots. The VLs correspond to values of $V_{SD}$ at which the chemical potential of differently charged ground states of the larger dot, $\mu_G^L(N_L)$, align with (in our case) the chemical potential of the *one* electron ground state of the smaller dot, $\mu_G^S(N_S = 1)$. Precisely speaking, such lines constitute a conductance threshold when the common chemical potential, $\mu_G^L(N_L) = \mu_G^S(1)$, lies between the source and drain Fermi energies, $E_F^S$ and $E_F^D$, *and* when $\mu_G^L(N_L +1) > E_F^S$. By contrast, (see Fig. 2) the *slanted* line segments FABC is characterized by alignment of $\mu_G^L(1)$ with the source Fermi energy, $E_F^S$. At point A: $\mu_G^S(1) >> \mu_G^L(1)$; hence, to reach the drain, electrons must pass through a high energy virtual state and current is un-observable. But at B, the triple alignment, $\mu_G^S(1) = \mu_G^L(1) = E_F^S$, occurs, and elastic current turns on. The VL from B to B' remains a current threshold until, at B', $\mu_G^L(2)$ enters the bias window (i.e. at B', $\mu_G^L(2) = E_F^S$). Segment BC extends segment AB (hence $\mu_G^L(1) = E_F^S$) but to the right of B we have $\mu_G^S(1) < \mu_G^L(1)$. This implies that an electron transiting from large to small dot must lose energy, presumably via coupling with bosons.

Next, the line through D and B' is characterized by degeneracy of $\mu_G^L(2)$ and $E_F^S$. At point D we have $\mu_G^S(1) = \mu_G^L(2) = E_F^S$, and the VL extending upward from D is the second current threshold. This pattern is repeated until finally at point E six electrons are trapped in the larger dot before the $N_S=1$ state aligns with the common Fermi energies of the leads.

As noted above, each VL pairs with a "kink" in negative $V_{SD}$. The main significance of these "kinks" here is that they represent displacements of the $V_{SD}$ conductance threshold which occur as additional electrons become trapped in the larger dot. This displacement results from the electrostatic influence of the added electron on the small dot and therefore provides a measure of the inter-dot Coulomb coupling, which we can thereby estimate as $E_{electro} \approx 1$ meV [13].

Summarizing, the $N_S=1$ state of the smaller dot serves to spectroscopically probe the larger dot. The final visible VL is B-B', giving a total of six discernable states of the larger dot. We now relate the VL spacing to the electronic structure of the small dot employing a modification of the CI model [16] in conjunction with an assumed F-D spectrum [17] for the single-particle levels (see Fig. 3c). In the standard CI model, the energy to add an electron to the dot, which is the difference between two neighboring chemical potentials, is decomposed into a "charging energy", $E_C$, and a single-particle level spacing, i.e. $\mu_G^L(N+1) - \mu_G^L(N) = E_C + \Delta$, where $\Delta$ is the spacing between the single-particle levels at the Fermi surface. The F-D spectrum for a circular parabolic potential in a B-field expresses each state with two quantum numbers, $E_{n,l} = (2n + |l| + 1)\hbar\overline{\omega} - l\omega_c$, where $\overline{\omega} \equiv \sqrt{\omega_0^2 + \omega_c^2}$ and where $\omega_0$ is the bare confining potential frequency and $\omega_c$ is the cyclotron frequency. The range of $n$ is $[0,\infty]$ and the range of $l$ is $[-\infty,\infty]$, giving the radial quantum number and the z-component of angular momentum, respectively (we also assume spin degeneracy of each state) [18]. The charging energy is usually expressed in terms of a constant dot capacitance; however, microscopically, it is in fact a combination of direct and exchange Coulomb integrals between occupied states and thus is not constant (i.e. it depends on which states are occupied). We make the simplifying assumption here that $E_C$ is constant *within a shell*. Thus, $E_C \equiv E_{CS}$ when adding the second s electron to the first shell; $E_C \equiv E_{C-SP}$ when adding the *first* p electron to the doubly occupied s shell; and $E_C \equiv E_{CP}$ when adding p electrons to the partially filled p shell. Further, we assume the dot is azimuthally symmetric, whereupon $E_{0,1} = E_{0,-1}$ at zero B-field (which gives fourfold degeneracy when spin is included).

We assume a voltage drop along the current path proportional to distance (allowing a conversion of $V_{SD}$ differences to energies) Since the s shell is (spin) degenerate, $\mu_G^L(2) - \mu_G^L(1) = E_{CS} = \Delta V_{SD}/3$, i.e. there is no single particle level spacing. Here, $\Delta V_{SD}$ is the spacing between the VLs of the two s states. The fourfold degeneracy of the p shell implies, similarly, that $E_{CP} = \Delta V_{SD}/3$, where now $\Delta V_{SD}$ is the VL spacing between any neighboring pair of p states. The spacing between the second VL of the s shell and the first one of the p shell contains both a level spacing, $E_{0,1} - E_{0,0}$, given by the confinement energy $\hbar\omega_0$ and the charging energy, $E_{C-SP}$. $E_{CS}$ and the average $E_{CP}$ are estimated to be 12 meV and 5 meV, respectively. Theoretical and experimental estimates of $E_{CS}$, $E_{CP}$ and $E_{C-SP}$ exist in the literature and typical values are 30 meV, 20 meV and 11 meV (resp.) theoretically [19] and 23 meV, 18 meV and 7 meV, experimentally [7]. Our dot size (diameter) is more than twice those discussed in previous reports, so the agreement of our results with earlier work is reasonable. Note that $E_{C-SP}$ cannot be derived from our measurements since we do not, *a prior*, know $\hbar\omega_0$, but projections of the above theoretical and experimental data to our dot size (combined with our own values of $E_{CS}$ and $E_{CP}$) leads us to estimate $E_{C-SP} \approx 5$ meV. This leads to an estimate of $\hbar\omega_0 \approx 15$ meV. In addition, we estimate an energy offset between the one electron states of the larger and smaller dots of $\approx 45$ meV simply from the $V_{SD}$ value at point B (with the aforementioned assumption of proportional voltage drop). The CI numerical analysis shown in the upper right inset of Fig. 2 is performed using all of these derived energies. The agreement seems quite satisfactory.

In order to estimate the energy offset and confinement energy directly, we have performed single dot spectroscopy using a micro-photoluminescence technique and 10-20 meV for the confinement energy and 50 meV for the energy offset are obtained [20]. These values are in good agreement with those from the transport measurements.

The evolution of the position of the VLs with B-field provides information on the angular momenta of the orbitals. Figures 3(a) and (b) show the B-field dependences of the position of the s shell VLs ($S_1$ and $S_2$), and that of p shell VLs ($P_1$ to $P_4$), respectively. The VLs shift in pairs with increasing B-field, indicating the presence of spin-paired orbitals. $S_1$ and $S_2$ suffer only a small diamagnetic shift with B-field, as expected. Their relative position, as indicated by $V_{SD}$ of the VLs, is very weakly dependent on B-field, suggesting a possible contribution from Zeeman energy. Two different types of evolution, $P_1$-$P_2$ or $P_3$-$P_4$, can be recognized in the p shell. This indicates that $P_1$-$P_2$ and $P_3$-$P_4$ form spin-paired states. Figure 3(c) shows calculation of F-D spectrum with $\hbar\omega_0 = 15$ meV. $P_1$ and $P_2$ recapitulate the behavior of $E_{0,1}$. On the other hand, the orbital resulting in the $P_3$ and $P_4$ lines experience a crossing which, according to the F-D spectrum, occurs near 6 T as indicated by arrow in Fig. 3(c). Only weak evidence of this level crossing is observed, and that occurs near 2 T. This departure from the F-D spectrum is probably due to non-parabolicity and deformation potential shape of the dot [21].

In conclusion, we have demonstrated single dot spectroscopy in a single pair of InAs coupled dots with precise manipulation of the energy levels of the dots. By analyzing the VLs in the non-linear transport regime for $V_G$ below the onset of Coulomb oscillations, the electronic shells of the single dot are studied up to the p shell. A calculation using the estimated energies reproduces the experimental data well. In the B-field dependent measurements, spin-paired states following F-D spectrum are observed.

The authors thank T. Nakaoka and Y. Arakawa for measurements of micro-photoluminescence and T. Inoshita, T. Sato and K. Yamada for valuable discussions. The authors acknowledge financial supports from the DARPA grant number DAAD19-01-1-0659 of the QuIST program, SORST-JST, the Grant-in-Aid for Scientific Research A (No. 40302799) and Focused Research and Development Project for the Realization of the World's

*Corresponding author. Present address:
Tarucha Mesoscopic Correlation Project, NTT Atsugi Research and Development Center, 4S-308S, 3-1, Wakamiya, Morinosato, Atsugi-shi, Kanagawa, 243-0198, Japan
*E-mail address:* ta-ota@tarucha.jst.go.jp Tel: +81-46-248-4000, Fax: +81-46-248-4014

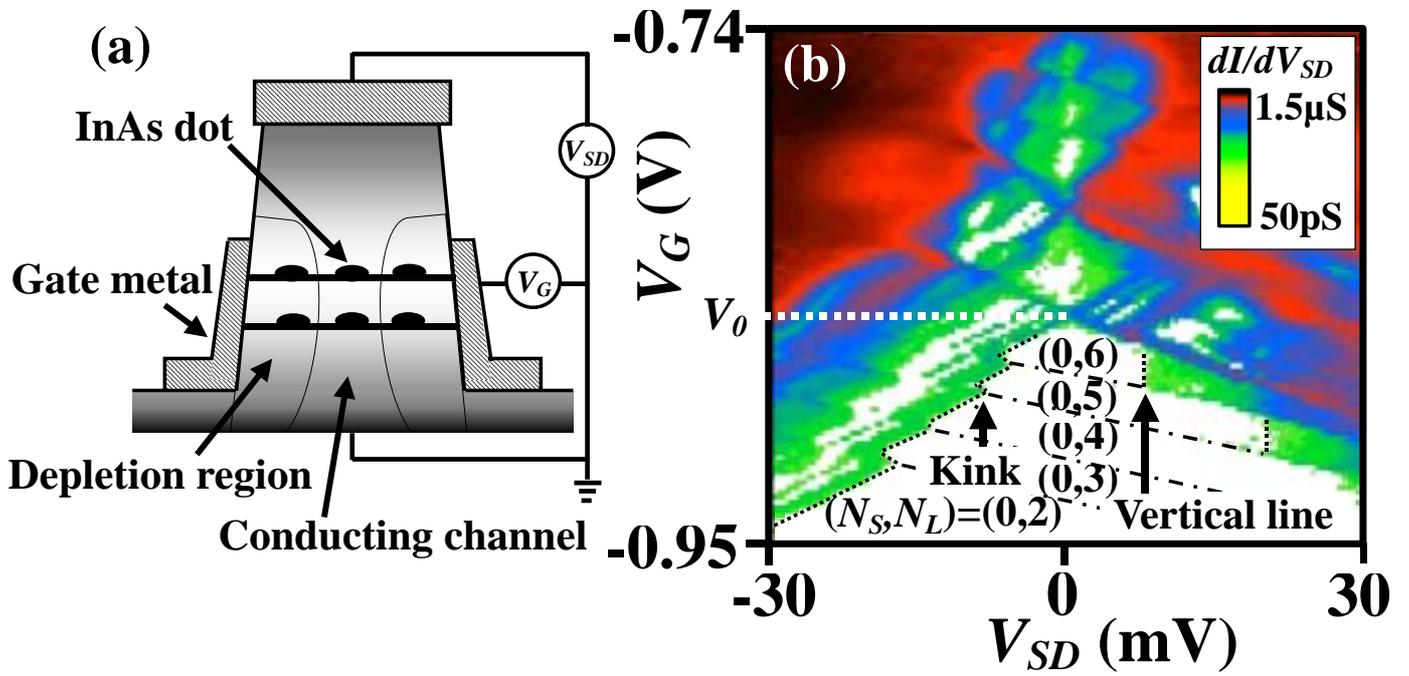

**Fig. 1**

(a) Schematic illustration of cross-section of the sub-micron size single-electron transistor used for the experiment and (b) color log-scale plot of $dI/dV$ as a function of $V_{SD}$ and $V_G$ close to $V_{SD}$ =0 V.

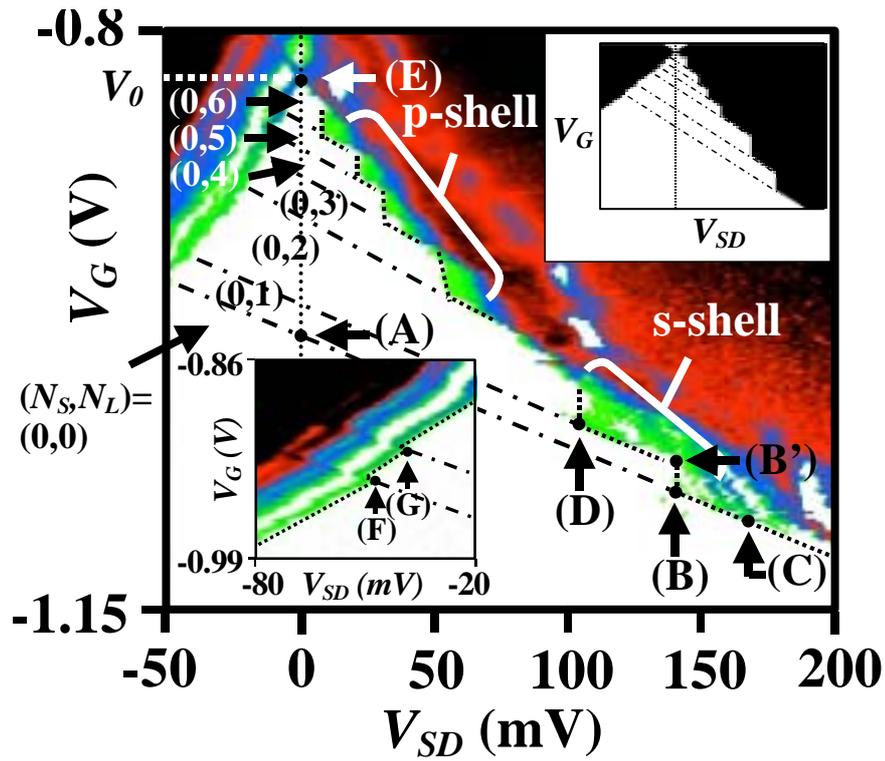
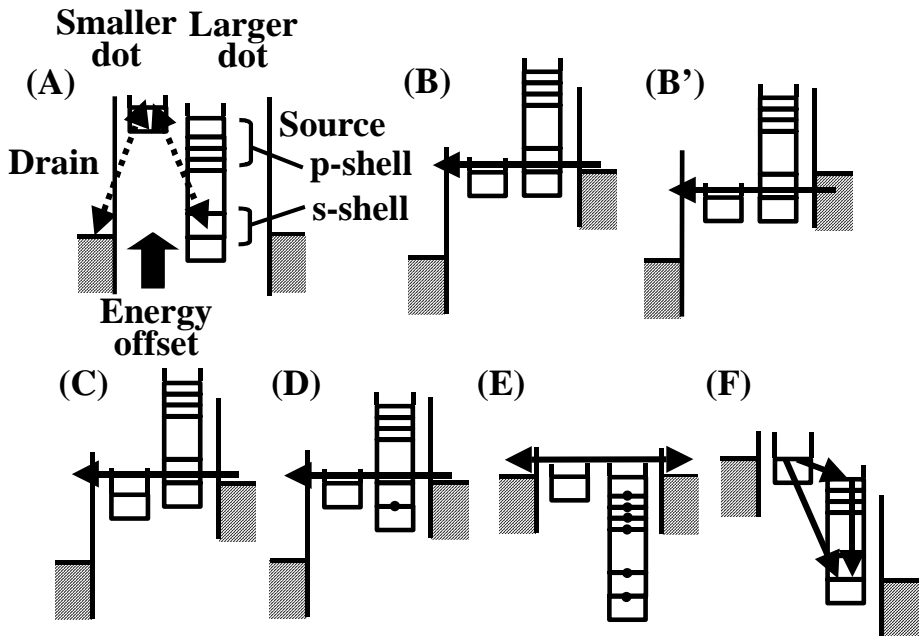

**Fig. 2**

Color log-scale plot of *dI/dV* as a function of $V_{SD}$ and $V_G$ in wider $V_{SD}$ range and configurations of energy-level diagram corresponding to points (A) to (F). The inset is calculated *dI/dV* plot using a simple constant interaction model.

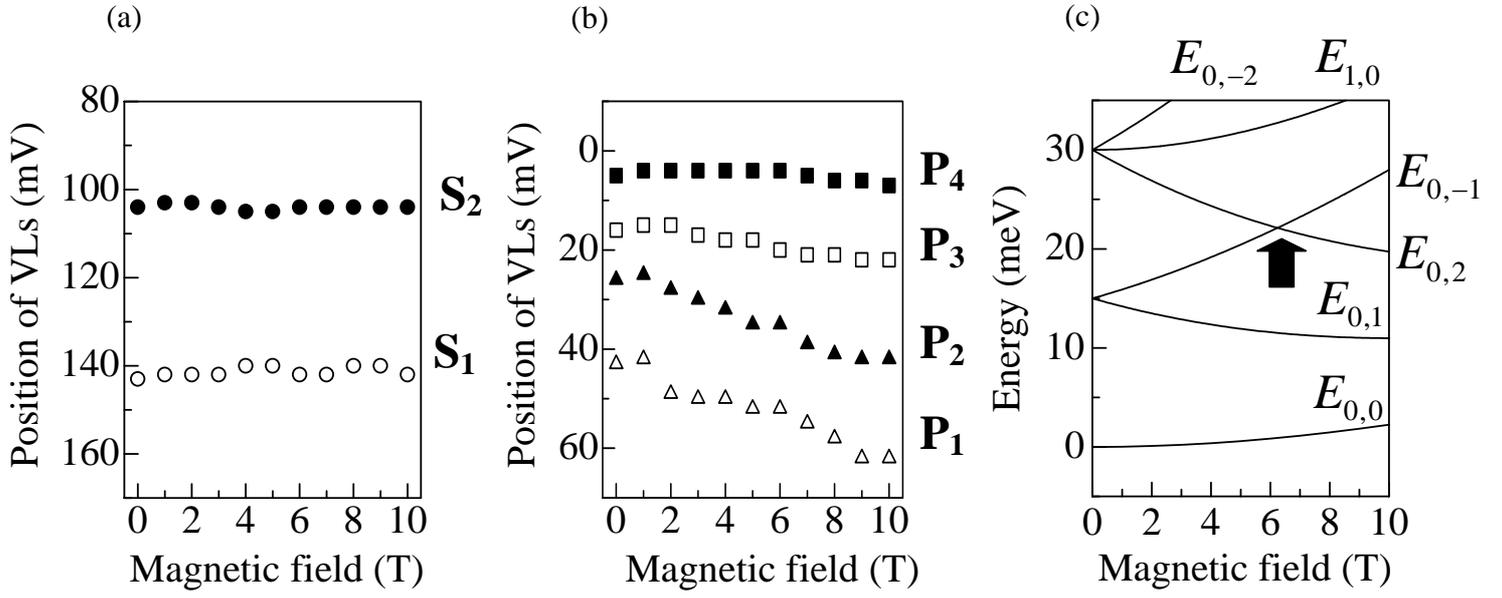

**Fig.3**

(a) The magnetic field dependence of the position of the s shell VLs ($S_1$ and $S_2$), (b) that of p shell VLs ($P_1$ to $P_4$) and (c) Fock-Darwin states with $\hbar\omega_0$ =15 meV.